\documentclass[referee]{mn2e}
\usepackage{graphics}

\makeatletter

\providecommand{\LyX}{L\kern-.1667em\lower.25em\hbox{Y}\kern-.125emX\@}

\newcommand{\beq}{\begin{equation}}
\newcommand{\eeq}{\end{equation}}
\newcommand{\bea}{\begin{eqnarray}}
\newcommand{\eea}{\end{eqnarray}}

\def\laq{~\raise 0.4ex\hbox{$<$}\kern -0.8em\lower 0.62
ex\hbox{$\sim$}~}
\def\gaq{~\raise 0.4ex\hbox{$>$}\kern -0.7em\lower 0.62
ex\hbox{$\sim$}~}

\makeatother
\begin{document}

\title{Acceleration at \( z>1 \)?}

\author[Luca Amendola]{Luca Amendola\\
INAF - Osservatorio Astronomico di Roma, Via Frascati 33, 00040 Monte
Porzio Catone (Roma) Italy, \\ amendola@mporzio.astro.it}
\date{Accepted ....,
      Received }
\pagerange{\pageref{firstpage}--\pageref{lastpage}}
\maketitle
\label{firstpage}\date{\today }

\begin{abstract}
Most models of dark energy predict the beginning of the accelerated
epoch at \( z\leq 1 \). However, there are no observational or theoretical
evidences in favor of such a recent start of the cosmic acceleration.
In fact, a model of dark energy coupled to dark matter is explicitely
constructed that \emph{a}) is accelerated even at high \( z \); \emph{b})
allows structure formation during acceleration; and \emph{c}) is consistent
with the type Ia supernovae Hubble diagram, including the farthest
known supernova SN1997ff at \( z\approx 1.7 \). It is shown that
the accelerated epoch in this model could have started as early as
\( z\approx 5 \).
\end{abstract}


\section{Introduction}

Cosmic acceleration is one of the most exciting discoveries of the
recent years. The combined data from cosmic microwave background (CMB,
see Netterfield et al. 2001, Lee et al. 2001, Halverson et al. 2001),
cluster masses and abundance (see e.g. Bahcall et al. 2002) and supernovae
Type Ia (SNIa) Hubble diagrams (Riess et al. 1998, Perlmutter et al.
1999) indicate that the universe is currently dominated by a very
weakly clustered component that is able to accelerate the expansion.
This component, denoted dark energy or quintessence, is supposed to
fill roughly 70\% of the cosmic medium.

However, the nature of the dark energy is still enigmatic (Wetterich
1988; Ratra \& Peebles 1988; Frieman et al. 1995; Caldwell et al.
1998). Its equation of state and its interaction with dark matter
are in fact so far subject only to very weak constraints (Perlmutter
et al. 1999, Huey et al. 1999, Baccigalupi et al. 2002, Corasaniti
\& Copeland 2002, Bean \& Melchiorri 2002, Amendola et al. 2002) so
that there exist a large variety of different dark energy models still
viable (see for instance the review by Peebles \& Ratra 2002). Perhaps
the only characteristic common to almost all proposed models of dark
energy is that its domination started very recently: the present epoch
of acceleration was preceded by a decelerated epoch at redshift \( z\geq 1 \)
in which structure formed. Two exceptions to this scheme are the model
of Dodelson et al. 2000, in which the dark energy is periodically dominating, and the
generic quintessence model of Lee \& Ng 2002, in which the dark energy
was dominating again at very large redshifts; in both cases the expansion
was however decelerated at \(z \) of order unity. 
The fact that the acceleration sets in just recently in the cosmic
history is one aspect of the {}``coincidence problem{}'' that demands
an explanation: see for instance the discussion based on the anthropic
principle by Vilenkin (2001). Another aspect of the {}``coincidence
problem{}'' is commonly phrased as why the dark energy and the dark
matter densities happen to be similar just today (Zlatev et al. 1999). 

The existence of a decelerated epoch rests on three arguments, one
theoretical and two observational. The theoretical argument is based
on the most common models of dark energy. In fact, a FRW universe
with scale factor \( a=(1+z)^{-1} \) filled with pressureless dark
matter (subscript \( m \)) and non-interacting dark energy (subscript
\( \phi  \)) with a constant equation of state \( w_{\phi }=1+p_{\phi }/\rho _{\phi } \)
is decelerated before the epoch \begin{equation}
\label{zacc}
z_{acc}=[(2-3w_{\phi })(\Omega _{\phi }/\Omega _{m})]^{1/(3-3w_{\phi })}-1.
\end{equation}
 Here and in the following the density parameters \( \Omega _{i} \)
refer to the present quantity of the \( i \)-th component. Given
the current estimates \( \Omega _{m}\approx 0.3\pm 0.1,\Omega _{\phi }=0.7\pm 0.1 \)
one has \( z_{acc}\leq 1 \) for all values of \( w \). Even allowing
for a large curvature, e.g. \( |\Omega _{k}|\equiv |1-(\Omega _{m}+\Omega _{\phi })|<0.3 \),
more than three sigma away from CMB measurements (de Bernardis et
al. 2001), it is not possible to go beyond \( z_{acc}\approx 1.3 \).
Therefore, an hypothetical observation of a value of \( z_{acc} \)
significantly larger than unity would signal that some of the assumptions
leading to (\ref{zacc}) are false.

The two observational arguments are as follows. First, since gravitational
instability is ineffective in an accelerated regime, it seems that
an extended accelerated era is in contrast with the observed large
scale structure. Second, the recent supernova SN1997ff at \( z\approx 1.7 \)
is consistent with a decelerated expansion at the epoch of light emission
(Benitez et al. 2001, Riess et al. 2002, Turner \& Riess 2001) and
seems to provide {}``a glimpse of the epoch of deceleration{}''
(Benitez et al. 2001).

The aim of this paper is to show that all three arguments are not
generally true: they are in fact valid only in a restricted class
of models. As a counterexample, a simple flat-space model with a constant
equation of state will be explicitely constructed that can be accelerated
at large \( z \), allows structure formation and is not in conflict
with the current supernova data, including SN1997ff. Such a dark energy
model, based on a coupling of dark energy to dark matter (Amendola
2000), makes very strong and unique predictions that can be easily
tested in the near future. 

Some of the results here reported have been already applied to a specific
model of string-inspired dark energy (Amendola et al. 2002). This
paper generalizes the arguments without referring to specific realizations
and performs a likelihood comparison to the SNIa data.

\section{The \protect\( z\approx 1.7\protect \) supernova}

Let us start with the last argument, i.e. the constraints imposed
by the farthest type Ia supernova known so far, SN1997ff (Benitez
et al. 2001). The recent assessment of the lensing magnfication increased
the apparent magnitude of SN1997ff by \( 0.34\pm 0.12 \) mag (Riess
et al. 2002). The result is conveniently expressed in terms of {}``Milne's
deviation{}'', i.e. the deviation \( \Delta (m-M) \) of the distance
modulus \( m-M \) from that of a Milne model (i.e. a hyperbolic empty
universe, \( \Omega _{m}=\Omega _{\Lambda }=0 \) which has a constant
expansion velocity). In Benitez et al. (2001) and Riess et al. (2002)
it was found \( \Delta (m-M)\approx -0.15\pm0 .34 \) at \( z\approx 1.755 \),
 in good agreement with a best fit flat \( \Omega _{m}=0.35,\Omega _{\Lambda }=0.65 \)
cosmology. As we anticipated, for such a cosmology the redshift \( z=1.755 \)
is indeed already well within the decelerated epoch, which started
at \( z_{acc}=0.548 \). However, this does not mean that all models
which are accelerated at large \( z \) are ruled out (even without
mentioning the still unclear experimental uncertainties of the supernova
detection). In fact, let us calculate the luminosity distance in Milne's
model. The luminosity distance in a FRW metric with scale factor \( a=(1+z)^{-1} \)
is defined as\begin{equation}
d_{L}(z)=\frac{1+z}{H_{0}}S\left[ \int _{0}^{z}\frac{dz'}{E(z')}\right] ,
\end{equation}
where the function \( S[x] \) is \( |\Omega _{k}|^{-1/2}\sin (|\Omega _{k}|^{1/2}x),x,|\Omega _{k}|^{-1/2}\sinh (|\Omega _{k}|^{1/2}x) \)
depending on the spatial curvature \( k=1,0,-1 \), respectively,
and where the function \( E(z) \) is derived from the Friedmann equation
written as\begin{equation}
H(z)=H_{0}E(z).
\end{equation}
 For instance, in the case of a model with matter (subscript \( m \)),
dark energy (subscript \( \phi  \)) with constant equation of state
\( w_{\phi } \) and curvature, one has\begin{equation}
\label{standard}
E^{2}(z)=\Omega _{m}(1+z)^{3}+\Omega _{\phi }(1+z)^{3w_{\phi }}+\Omega _{k}(1+z)^{2},
\end{equation}
 where \( \Omega _{k}=1-\Omega _{m}-\Omega _{\phi }. \) In an empty
hyperbolic universe we have then \( \Omega _{k}=1 \) and\begin{equation}
E(z)=(1+z),
\end{equation}
so that the luminosity distance is\begin{equation}
\label{milne}
d_{L(Milne)}=\frac{(1+z)}{H_{0}}S\left[ \int _{0}^{z}\frac{dz'}{E(z')}\right] =\frac{z^{2}+2z}{2H_{0}}.
\end{equation}
It appears then that \( d_{L(Milne)}\sim z^{2} \) for large redshifts.
Any model with a luminosity distance that grows slower than \( z^{2} \)
yields \( \Delta (m-M)<0 \) (as SN1997ff) beyond some \( z \) .

Perhaps the simplest model one can construct with such a property
is a flat space with a single fluid with constant equation of state
\( w_{e} \). In this case one has in fact\begin{equation}
\label{single}
H^{2}=H_{0}^{2}a^{-3w_{e}},
\end{equation}
and the luminosity distance (for \( w\not =2/3) \)\begin{equation}
\label{dl}
d_{L}=(1+z)\int _{0}^{z}\frac{dz'}{H(z')}=\frac{2(1+z)}{(2-3w_{e})H_{0}}\left[ (1+z)^{-\frac{3w_{e}}{2}+1}-1\right] .
\end{equation}
For \( w_{e}<2/3 \) and large \( z \) we have \( d_{L}\sim z^{2-3w_{e}/2} \).
It follows that, at large \( z \) and for any \( w_{e}>0 \), Milne's
model gives always larger distances (and therefore fainter luminosity,
i.e. larger apparent magnitudes) than the single-fluid flat-space
evolution. In Fig. 1 I plot \( \Delta (m-M) \) for various values
of \( w_{e} \), along with the reference flat model \( \Omega _{m}=0.35,\Omega _{\Lambda }=0.65 \).
It appears clearly that values of \( w_{e} \) around \( 0.4\pm 0.1 \)
are acceptable, as it will be confirmed by the likelihood analysis
below.

\begin{figure}
{\centering \includegraphics{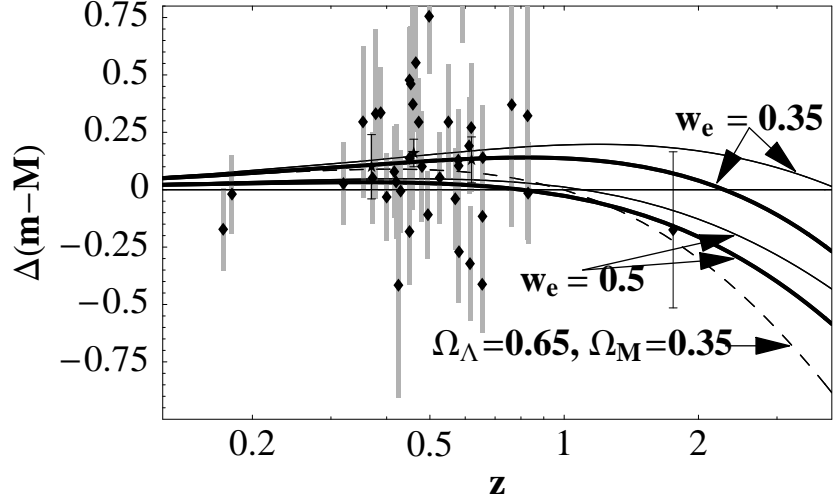} \par}

\caption{The thin curves represent the quantity \protect\( \Delta (m-M)\protect \)
for two values of \protect\( w_{e}\protect \). The thick curves include
a component \protect\( \Omega _{b}=0.05\protect \) of uncoupled baryons.
The dashed curve is the reference model. The gray errorbars represents
the high-redshift data of Perlmutter et al. (1999), while the black
errorbars summarize all the supernova data at large \protect\( z\protect \)
derived from Benitez et al. (2001).}
\end{figure}

However, according to the observations, our universe is composed by
a mixture of, say, 30\% of clustered matter (which includes a 5\%
of baryons) and 70\% unclustered dark energy. Therefore the single-fluid
model (\ref{single}) can account for the observations only if both
components scale with the same power \( w_{e} \). This is exactly
what happens assuming a direct interaction between dark matter and
dark energy in the regime of strong coupling (Amendola 2000, Amendola
\& Tocchini-Valentini 2001, Gasperini et al. 2001). Let us assume
infact two fluids: one (subscript \( m \)), with equation of state
\( w_{m} \), is later to be identified with dark matter, and the
other (subscript \( \phi  \)), with equation of state \( w_{\phi } \),
to be identified with dark energy. Assume then a direct interaction
term \( \delta  \) that trasfers energy from one component to the
other. The conservation equations are\begin{eqnarray}
\dot{\rho }_{\phi }+3Hw_{\phi }\rho _{\phi } & = & -\delta ,\\
\dot{\rho }_{m}+3Hw_{m}\rho _{m} & = & \delta ,
\end{eqnarray}
where the Friedmann equation reads\begin{equation}
3H^{2}=\kappa ^{2}\left( \rho _{\phi }+\rho _{m}\right) ,
\end{equation}
and \( \kappa ^{2}=8\pi G \) . It appears then that a solution with
\( \rho _{m}\sim \rho _{\phi } \) exists if (Zimdahl et al. 2001,
Amendola \& Tocchini-Valentini 2002)\begin{equation}
\delta =3H\Omega _{\phi }\left( w_{m}-w_{\phi }\right) \rho _{m}.
\end{equation}
Therefore, the direct interaction induces on both components an effective
equation of state \begin{equation}
w_{e}=w_{m}+\Omega _{\phi }\left( w_{\phi }-w_{m}\right) ,
\end{equation}
such that \begin{equation}
\rho _{m}\sim \rho _{\phi }\sim a^{-3w_{e}}.
\end{equation}
The Friedmann equation becomes then \( H^{2}=H_{0}^{2}a^{-3w_{e}} \)
as assumed in (\ref{single}). 

Accelerated cosmological solutions with \( \rho _{m}\sim \rho _{\phi } \),
denoted \emph{stationary} solutions because \( d(\rho _{m}/\rho _{\phi })/dt=0 \),
has been studied in several papers (see e.g. Amendola 1999, Chimento
et al. 2000, Amendola \& Tocchini-Valentini 2001, Gasperini et al.
2001, Pietroni 2002, Bonanno \& Reuter 2002, Gromov et al. 2002) even before the evidences
for acceleration (Carvalho et al. 1992, Wetterich 1995). This class
of dark energy models has been invoked to explain the coincidence
problem: in fact, in such models the dark matter and the dark energy
scale identically with time, and as a consequence their ratio is frozen
to the observed value from some time onward. It is indeed not difficult
to produce accelerated expansions in which the ratio \( \rho _{m}/\rho _{\phi } \)
is of order unity at all times after equivalence (Amendola \& Tocchini-Valentini
2001). 

In Fig. 2 we plot the contours of the likelihood function obtained
analysing the SNIa data of Perlmutter et al. (1999) to determine the
allowed range of \( w_{e} \) (we consider the 38 high-\( z \) SNe
plus the 16 low-\( z \) SNe of their fit C, plus the supernova at
\( z=1.7 \)). In the upper part of the plot we show the likelihood
\( L(\Omega _{m},w_{\phi }) \) for a {}``standard{}'' flat-space
model composed of two uncoupled fluid as in Eq. (\ref{standard})
with \( \Omega _{k}=0 \): a fraction \( \Omega _{m} \) of matter
with equation of state \( w_{m}=1 \) and a fraction \( \Omega _{\phi }=1-\Omega _{m} \)
of dark energy with constant equation of state \( w_{\phi } \). This
plot reproduces the similar one in Perlmutter et al. (1999) (see also
Dalal et al. 2001 for a further generalization) and has been obtained
by marginalizing over the {}``nuisance{}'' parameters  \( M \) (the
supernovae absolute magnitude) and \( \alpha  \) (the slope of the
stretch factor relation, see Perlmutter et al. 1999 for the definitions
of these parameters). In the limit of \( \Omega _{m}\to 0 \) the
luminosity distance becomes as in the stationary case (\ref{dl})
with \( w_{\phi }=w_{e} \). Therefore, in the lower panel we show
the one-dimensional likelihood \( L(\Omega _{m}=0,w_{\phi }) \),
corresponding to the section at \( \Omega _{m}=0 \) of the upper
panel. As we will show in the following, it is necessary to decouple
the baryons from the dark energy, so that the simple stationary case
(\ref{dl}) has to be modified with the addition of a fraction \( \Omega _{b} \)
of uncoupled baryons. The likelihood \( L(\Omega _{b}=0.05,w_{\phi }) \)
of this case is shown as a thin line. Finally, the intermediate likelihood
drawn in dashed line includes also the supernova SN1997ff. The conclusion
is that \begin{equation}
w_{e}\in (0.2,0.6)
\end{equation}
 at 95\% c.l. (\( w_{e}=0.4\pm 0.1 \) at one sigma), with minimal
variations including baryons and including SN1997ff. Notice that this
is a one-parameter fit to the SNIa, with a \( \chi ^{2}/ \)d.o.f.\( =1.001 \)
for the best fit, just as good as the reference pure-\( \Lambda  \)
model.
\begin{figure}
{\centering \resizebox*{!}{12cm}{\includegraphics{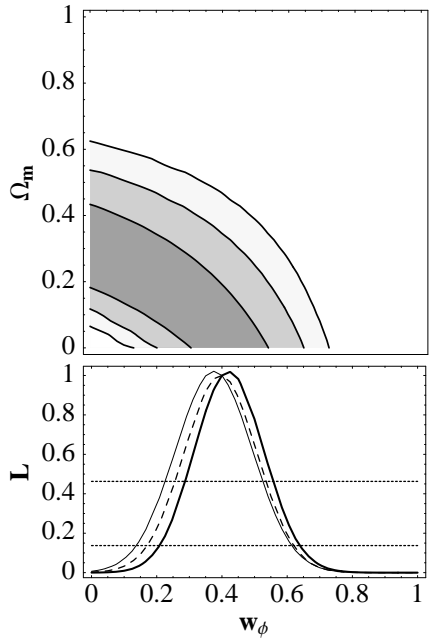}} \par}

\caption{Upper panel: marginalized likelihood function for a flat space model
with two fluids, one with equation fo state \protect\( w_{m}=1\protect \)
and the other \protect\( w_{\phi }=const.\protect \) The contours
enclose 67\%, 90\% and 95\%, dark to light gray. Lower panel: likelihood
functions for the equation of state for a stationary model. Thick
line: without baryons; thin line: including a fraction \protect\( \Omega _{b}=0.05\protect \)
of uncoupled baryons; dashed line: including baryons and SN1997ff.
The two horizontal lines mark the levels corresponding to 67\% and
95\% c.l.}
\end{figure}

Naturally, the stationary solution is acceptable only if one can show
that the component \( \rho _{m} \) clusters while the dark energy
component \( \rho _{\phi } \) does not. This is the argument of the
next section.

\section{Growth of perturbations}

Let us now identify the dark energy component with a scalar field.
It has been shown in Amendola \& Tocchini-Valentini (2001) that, assuming
\( w_{\phi } \) and \( w_{m} \) constant, the only scalar field
model that reduces to the equations above requires a coupling \begin{equation}
\label{lincoup}
\delta =\sqrt{2/3}\kappa \beta |\dot{\phi }|\rho _{m},
\end{equation}
 with \begin{equation}
\label{etabeta}
\beta =-\frac{3(w_{m}-1+\Omega _{\phi }(w_{\phi }-w_{m}))}{\sqrt{2(w_{m}-1+\Omega _{\phi }(1-w_{m}+w_{\phi })}}=\frac{3\Omega _{\phi }}{\sqrt{2w_{\phi }\Omega _{\phi }}}\left( 1-w_{\phi }\right) ,
\end{equation}
 where the second equality holds for pressureless dark matter, \( w_{m}=1 \),
which is the case we study below. Moreover, it is easy to show that
\( w_{\phi }= \)const. implies an exponential potential \( U=U_{0}e^{-\sqrt{2/3}\mu \kappa \phi } \)
where \begin{equation}
\label{mueta}
\mu =\frac{3(w_{m}+\Omega _{\phi }(w_{\phi }-w_{m}))}{\sqrt{2(w_{m}-1+\Omega _{\phi }(1-w_{m}+w_{\phi })}}=\frac{3}{\sqrt{2w_{\phi }\Omega _{\phi }}}\left[ 1+\Omega _{\phi }(w_{\phi }-1)\right] 
\end{equation}
(again the second equality is for \( w_{m}=1 \) ). The equations
for the matter (from now on \( w_{m}=1 \)) and the scalar field component
are then\begin{eqnarray}
\ddot{\phi }+3H\dot{\phi }+U_{,\phi } & = & -\sqrt{2/3}\kappa \beta \rho _{m},\\
\dot{\rho }_{m}+3H\rho _{m} & = & \sqrt{2/3}\kappa \beta \rho _{m}\dot{\phi }.\label{sys} 
\end{eqnarray}
For \( \beta >\sqrt{3}/2 \) and \( \mu >(-\beta +\sqrt{18+\beta ^{2}})/2 \)
(Amendola 2000), the system converges toward a global attractor with
a power law expansion \( a(t)\sim t^{p} \) , characterized by\begin{eqnarray}
\Omega _{\phi } & = & \frac{4\beta ^{2}+4\beta \mu +18}{4(\beta +\mu )^{2}},\\
w_{e} & = & \frac{\mu }{\mu +\beta },\\
p & = & \frac{2}{3}\left( 1+\frac{\beta }{\mu }\right) ,
\end{eqnarray}
 which is accelerated for \( \mu <2\beta . \) 

The growth of perturbations in the above model has been studied in
Amendola \& Tocchini-Valentini (2002); here we just state the results.
For scales smaller than the horizon the perturbation equation for
the matter density contrast \( \delta _{m} \) can be written as\begin{equation}
\label{deltacsimp}
\delta ''_{m}+\left( 2+\frac{H'}{H}+\frac{3\beta }{\beta +\mu }\right) \delta _{m}'-\frac{3}{2}\gamma \delta _{m}\Omega _{m}=0,
\end{equation}
 where primes denote derivation with respect to \( \log a \) and
\( \gamma \equiv 1+4\beta ^{2}/3. \) Employing the trace of the Einstein
equations \begin{equation}
\frac{H'}{H}=-\frac{1}{2}\left[ 3+\kappa ^{2}\left( \frac{1}{2}\phi '^{2}-\frac{U}{H^{2}}\right) \right] =-\frac{3}{2}w_{e},
\end{equation}
we find the solution \( \delta _{m}=a^{m_{\pm }} \) where\begin{equation}
\label{sol}
m_{\pm }=\frac{1}{4}\left[ -4+3w_{e}-\frac{6\beta }{\beta +\mu }\pm \Delta \right] ,
\end{equation}
where \( \Delta ^{2}=24\gamma \Omega _{m}+(-4+3w_{e}-\frac{6\beta }{\beta +\mu })^{2} \). 

The observational constraints \( \Omega _{\phi }=0.7\pm 0.1 \) and
\( w_{e}=0.4\pm 0.1 \) confine the parameters \( \beta ,\, \mu  \)
in the shadowed region in Fig. 3, where we also plot \( m_{+}(\beta ,\, \mu ) \)
. For instance, putting \( w_{e}=0.4 \), as suggested by the SNIa,
and \( \Omega _{m}=0.3 \) we obtain a growth as fast as \( m_{+}\approx 2 \),
which reduces to \( m_{+}\approx 1 \) if \( w_{e}=0.5 \). This shows
that the dark matter fluctuations can grow even during the accelerated
phase, due to the extra pull of the dark energy coupling. 

The \( k \)-th plane-wave component of the scalar field perturbation
\( \varphi =\kappa (\delta \phi )/\sqrt{6} \) grows as \begin{equation}
\label{phipert}
\varphi =\varphi _{0}\delta _{c}\Omega _{c}(H_{0}/k)^{2}a^{-3w_{e}+2},
\end{equation}
 where \begin{equation}
\varphi _{0}=\beta -\frac{9}{\mu -2\beta +(\beta +\mu )\Delta }.
\end{equation}
For subhorizon wavelengths \( \varphi  \) (which is proportional
to \( \delta \rho _{\phi }/\rho _{\phi } \) ) remains always much
smaller than \( \delta _{m} \). It is interesting that even the dark
energy fluctuations grow (as \( a^{m+2-3w_{e}} \)) during the accelerated
epoch. 

This demonstrates that the two components may be indeed identified
with clustered matter and unclustered dark energy. Naturally, we have
only proved that the perturbations do grow during acceleration, not
that the present level of dark matter fluctuations is compatible with
observations. This requires however the knowledge of the behavior
of the perturbation throughout the post-decoupling era, not only during
the final accelerated stage. In detailed models such as the string-inspired
one discussed in Amendola et al. (2002) it can be shown that it is
indeed possible to match the observed level of fluctuations.

As an aside, let us comments on the consequences of the above results
on an application of the anthropic principle. According to Vilenkin
(2002), that generalized previous analyses by Weinberg (1987), the
amount of dark energy cannot exceed too much the observed value because
otherwise the perturbation growth stops before galaxies can form,
thereby preventing life. In particular, the anthropic principle {}``predicts{}''
that \( \Omega _{\phi } \) has to be significanty smaller than unity, i.e. \( \rho_{\phi}
\approx \rho_m \).
However, if dark energy is coupled to dark matter, perturbations can
grow even when \( \Omega _{\phi } \) is dominating. For instance,
a universe with \( \Omega _{\phi }>0.999 \) (i.e. \( \rho _{\phi }>1000\rho _{m} \))
admits a growth rate \( m=1 \) if \( \beta  \) is very large (\( \beta >50 \)
in this case). It seems therefore that the anthropic principle is
unable to constrain  \( \Omega _{\phi } \) in the general case.

\begin{figure}
{\centering \resizebox*{!}{11cm}{\includegraphics{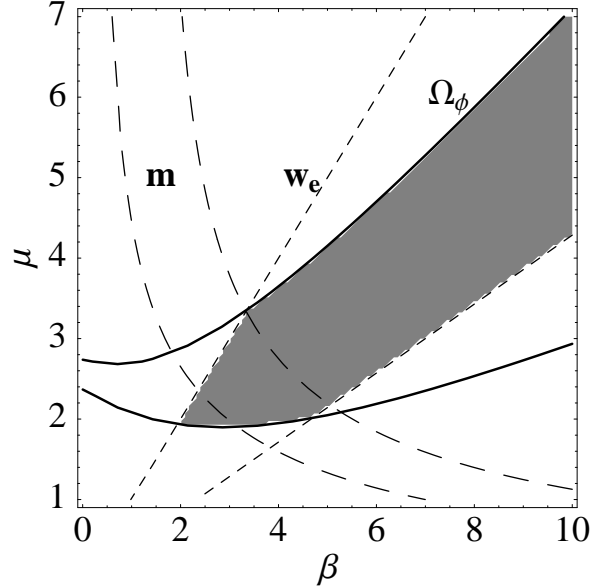}} \par}

\caption{The shadowed region represents the observational constraints on \protect\( \beta ,\mu \protect \).
Within the unbroken curves \protect\( \Omega _{\phi }\in (0.6,0.8)\protect \);
within the short-dashed curves \protect\( w_{e}\in (0.3,0.5)\protect \).
The long-dashed curves represent \protect\( m(\beta ,\, \mu )=1\protect \)
(left curve) and \protect\( m(\beta ,\, \mu )=2\protect \) (right
curve).}
\end{figure}

\section{When did the acceleration start?}

We have shown so far that the past duration of the present acceleration
is not limited by the supernovae data nor by the structure growth.
But how far in the past can we extend the accelerated stationary solution?
In this section we show that the limiting factor is to be found in
the present observed abundance of baryons.

In fact, the coupling between matter and dark energy is a {}``fifth
force{}'' comparable in strength to gravity. This can be seen already
from Eq. (\ref{deltacsimp}), in which the usual force term \( 3\delta _{m}\Omega _{m}/2=4\pi G(\delta \rho )_{m}/\rho _{crit} \)
is replaced by \( 3\gamma \delta _{m}\Omega _{m}/2=4\pi G\gamma (\delta \rho )_{m}/\rho _{crit} \).
In other words, to the lowest order, the effect of the dark matter/dark
energy interaction is to induce an effective gravitational constant
(Damour \& Nordvedt 1993) \begin{equation}
\tilde{G}=G\gamma =G\left( 1+\frac{4\beta ^{2}}{3}\right) .
\end{equation}
As long as the dark energy interacts only with dark matter, such a
fifth force is of course locally unobservable. However, if the dark
energy couples to baryons with strength \( \beta _{b} \) then the
interaction shows up to the post-post Newtonian (PPN) order as a non-zero
parameter \( \overline{\gamma } \) defined as in Hagiwara et al.
(2002). The relation between \( \overline{\gamma } \) and \( \beta _{b} \)
can be found by observing that the coupled system of dark energy and
matter fields \( \psi  \) can be derived from the Lagrangian\begin{equation}
L=\sqrt{|g|}(-\frac{R}{2\kappa ^{2}}+\frac{1}{2}\phi _{;\mu }\phi ^{;\mu }-V(\phi ))+L_{m}(\psi ,\tilde{g}_{\mu \nu }),
\end{equation}
 where the metric \( \tilde{g}_{\mu \nu }=e^{2a(\phi )}g_{\mu \nu } \)
is conformally related to \( g_{\mu \nu } \) by the factor \begin{equation}
e^{2a(\phi )}=e^{-2\kappa \sqrt{\frac{2}{3}}\beta _{b}\phi }=e^{-4\sqrt{\frac{2}{3}}\beta _{b}\hat{\phi }},
\end{equation}
and where the dimensionless field \( \hat{\phi }=\kappa \phi /2 \)
has been introduced to match the definition of Damour \& Esposito-Farese
(1992), Hagiwara et al. (2002) and Will (2002). Then, according to
Damour \& Esposito-Farese (1992) (see also Will 2002), the PPN parameter
\( \overline{\gamma } \) is defined as\begin{equation}
\overline{\gamma }=-\frac{2\alpha _{0}^{2}}{1+\alpha _{0}^{2}}=-\frac{4\beta _{b}^{2}}{3+2\beta _{b}^{2}},
\end{equation}
where \( \alpha _{0}=da(\phi )/d\phi  \). The upper limits on \( |\gamma | \)
are of the order of \( 10^{-4} \) (Hagiwara et al. 2002), so that
\( \beta _{b} \) is constrained to be smaller than \( 10^{-2} \)
roughly. For as concerns the cosmological evolution at small \( z \),
this is equivalent to assuming \( \beta _{b}=0 \), so that our starting
assumption that the dark energy couples only (or preferentially) to
dark matter, is justified. Similar species-dependent couplings have
been discussed in other contexts since the first proposal by Damour,
Gibbons, \& Gundlach (1990), see e.g. the astrophysical bounds discussed
by Gradwohl \& Frieman (1992).

However, if the baryons are uncoupled they dilute with the usual behavior
\( \rho _{b}\sim a^{-3} \) i.e. faster than the coupled dark energy/dark
matter fluid. This modifies the Friedmann equation (\ref{dl}) as
follows \begin{equation}
\label{bar}
H^{2}=H_{0}^{2}[(1-\Omega _{b})a^{-3w_{e}}+\Omega _{b}a^{-3}].
\end{equation}
There appears therefore an epoch in the past before which the baryons
were dominating and, as a consequence, the expansion decelerated.
Denoting with \( \Omega _{u} \) a generic uncoupled component (in
this case the baryons) this epoch \( z_{acc} \) in flat space is
\begin{equation}
\label{zaccb}
z_{acc}(\Omega _{u},w_{e})=[(2-3w_{e})(1-\Omega _{u})/\Omega _{u}]^{1/(3-3w_{e})}-1,
\end{equation}
which therefore replaces (\ref{zacc}). It appears then that the factor
that limits the acceleration epoch is the present abundance of baryons.
Assuming \( \Omega _{u}=\Omega _{b}\approx 0.05\pm0 .02 \) the maximum
\( z_{acc} \) turns out to be around 5 as shown in Fig. 4.

\begin{figure}
{\centering \resizebox*{!}{12cm}{\includegraphics{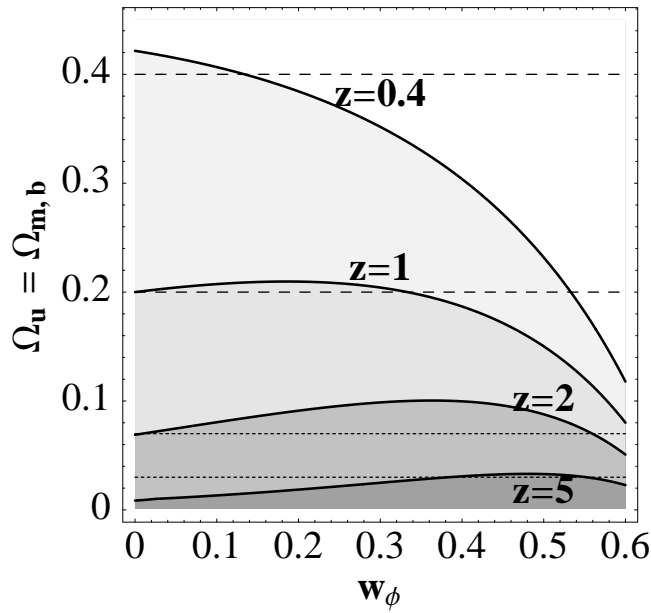}} \par}

\caption{Contour plots of the function \protect\( z_{acc}(\Omega _{u},w_{e})\protect \)
where the uncoupled component is either the total matter content \protect\( \Omega _{m}\protect \),
as in the standard dark energy models, or the baryon fraction \protect\( \Omega _{b}\protect \)
as in the model discussed in this paper. The observed total matter
fraction is within the dashed lines, while the baryon fraction is
within the dotted lines.}
\end{figure}

It is to be noticed that if \( \beta  \) and \( \mu  \) are constant
at all epochs, the present accelerated epoch is preceded by a decelerated
baryon-dominated epoch before \( z_{acc} \). Such an epoch would
however be in conflict with the CMB, as shown in Tocchini-Valentini
\& Amendola (2001), because of an extremely large integrated Sachs-Wolfe
effect on the CMB. It is therefore necessary to modify the simplest
case with, for instance, a modulation of the coupling parameter \( \beta  \)
or the potential \( U \) in order to prevent the baryon domination,
as in Amendola \& Tocchini-Valentini (2001) and in Amendola et al.
(2002). These models also account for the observed level of present
fluctuations.

\section{Conclusion}

This paper shows that high-\( z \) acceleration is a viable possibility
if dark energy couples to dark matter. Although the present abundance
of baryons limit the epoch of acceleration to \( z_{acc}<5 \) , this
is still much earlier than the standard models of uncoupled dark energy,
which hardly reaches \( z_{acc}=1 \). The future observations of
SNIa at high redshift will be well suited to detect or reject an early
acceleration.

A strong coupling between dark energy and dark matter can also be
detected through the biasing and the rate of growth of perturbations,
as discussed in Amendola \& Tocchini-Valentini (2002). In principle,
the coupling could be seen also in astrophysical objects like galaxy
clusters by comparing the forces felt by dark matter halos and baryons
(for instance, baryons in the intracluster gas), i.e. as an astrophysical
test of the violation of the equivalence principle. This approach,
discussed in detail in Gradwohl \& Frieman (1992), requires several
assumptions on the distribution of dark haloes. As emphasized by Peebles
(2002), however, it has the potentiality to open unexplored paths
in cosmology.

\section{References}
Amendola L. Phys. Rev. \textbf{D62}, 043511 (2000)\newline
Amendola L. and D. Tocchini-Valentini, Phys. Rev. \textbf{D64}, 043509
(2001)\newline
Amendola L. \& D. Tocchini-Valentini, astro-ph/0111535, Phys. Rev.
\textbf{D66}, 043528 (2002)\newline
Amendola L., C. Quercellini, D. Tocchini-Valentini and A. Pasqui,
(2002) astro-ph/0205097 \newline
Amendola L., M. Gasperini, D. Tocchini-Valentini and C. Ungarelli,
(2002) astro-ph/0208032\newline
Baccigalupi C., A. Balbi, S. Matarrese, F. Perrotta, N. Vittorio,
astro-ph/0109097, Phys.Rev. \textbf{D65} (2002) 063520\newline
Bahcall N. et al. 2002, astro-ph/0205490\newline
Bean R. \& Melchiorri A., astro-ph/0110472, Phys.Rev. \textbf{D65}
(2002) 041302\newline
Benitez N. et al., (2002) astro-ph/0207097\newline
 Bonanno A. \& M. Reuter (2002) Phys. Lett. B 527, 9\newline
Caldwell R.R., Dave R. \& Steinhardt P.J. (1998), Phys. Rev. Lett.
80, 1582 \newline
Carvalho J.C., J.A.S. Lima, I. Waga ,  Phys.Rev.\textbf{D46}:2404-2407,1992~\newline
Chimento L.P., A. S. Jakubi \& D. Pavon, Phys. Rev. \textbf{D62},
063508 (2000), astro-ph/0005070; \newline
Corasaniti P.S.\& E. Copeland, Phys.Rev. \textbf{D65} (2002) 043004\newline
Damour T.\& Esposito-Farese G., Class. Quantum Grav. 9, 2093 (1992)\newline
Damour T. \& Nordtvedt K., Phys. Rev. Lett. 70, 2217 (1993)\newline
Damour T., G. W. Gibbons and C. Gundlach, Phys. Rev. Lett., 64, 123,
(1990)\newline
Dalal N. et al., Phys. Rev. Lett., \textbf{87}, 141302 (2001)\newline
De Bernardis et al. Nature 404:955-959,2000\newline
 Dodelson S., M. Kaplinghat \& E. Stewart, Phys. Rev. Lett. 85, 5276
(2000)\newline
 Frieman J., C. T. Hill, A. Stebbins and I. Waga, Phys. Rev. Lett.
\textbf{75}, 2077 (1995).\newline
Gasperini M., F. Piazza and G. Veneziano, Phys. Rev. \textbf{D65},
023508 (2001). \newline
 Gradwohl B.A.\& Frieman, J. A. 1992 ApJ, 398, 407\newline
Gromov A., Yu. Baryshev \& P. Teerikorpi (2002) astro-ph/0209458\newline
Hagiwara K. et al., Phys. Rev. \textbf{D66} 010001-1 (2002), available
at pdg.lbl.gov\newline
Halverson N.W.\textit{et al.}, (2001) astro-ph/0104489. \newline
Huey G., Wang, L., Dave, R., Caldwell, R. R., \& Steinhardt, P. J.
(1999), Phys. Rev. D 59,063005. \newline
Lee A.T.\textit{et al.}, Ap. J. \textbf{561}, L1 (2001).\newline
 Lee W.\& K.-W. Ng astro-ph/0209093.\newline
Netterfield C.B.\textit{et al.}, astro-ph/0104460\newline
 Perlmutter S.\textit{et al}, Ap. J. \textbf{517}, 565 (1999).\newline
Peebles P.J.E.\& B. Ratra, astro-ph/0207347\newline
 Peebles P.J.E., astro-ph/0208037\newline
Pietroni M., hep-ph/0203085.\newline
Ratra B.\& P.J.E. Peebles, Phys. Rev. D37, 3406 (1988)\newline
Riess A.\textit{et al}, Astron. J. \textbf{116}, 1009 (1998).\newline
 Riess A. et al. Ap. J. \textbf{560}, 49 (2001)\newline
Tocchini-Valentini D. and L. Amendola, Phys.Rev. \textbf{D65}, 063508
(2002)\newline
Turner M. and A. Riess, astro-ph/0106051\newline
Vilenkin A., hep-th/0106083\newline
Weinberg S. Phys. Rev. Lett. 59, 2607 (1987)\newline
Wetterich C. (1988) Nucl. Phys. B., 302, 668; \newline
Wetterich C., Astronomy and Astrophys. \textbf{301}, 321 (1995).\newline
Will C., Living Rev. Relativity, 4 (2001), 4: cited on Sep. 19, 2002,
http://www.livingreview.org/Articles/Volume4/2001-4will/\newline
Zimdahl W., D. Pavon \& L. P. Chimento, Phys.Lett. B521 (2001) 133,
astro-ph/0105479 \newline
Zlatev I., L. Wang \& P.J. Steinhardt, Phys. Rev. Lett. 82 896 (1999).

\label{firstpage}
\end{document}